\documentclass[prl,twocolumn,superscriptaddress,showpacs,longbibliography]{revtex4-1}
\usepackage{mathrsfs}
\usepackage{amssymb, amsbsy, amsmath, latexsym, dsfont, array, layout,
graphicx,mathrsfs,color,ulem,bm}

\newcommand{\one}{\mathds{1}}
\newcommand{\ket}[1]{\left|{#1}\right\rangle}
\newcommand{\bra}[1]{\left\langle{#1}\right|}

\begin{document}

\title{Detecting topological invariants in non-unitary discrete-time quantum walks}
\author{Xiang Zhan}
\affiliation{Department of Physics, Southeast University, Nanjing 211189, China}
\author{Lei Xiao}
\affiliation{Department of Physics, Southeast University, Nanjing 211189, China}
\author{Zhihao Bian}
\affiliation{Department of Physics, Southeast University, Nanjing 211189, China}
\author{Kunkun Wang}
\affiliation{Department of Physics, Southeast University, Nanjing 211189, China}
\author{Xingze Qiu}
\affiliation{Key Laboratory of Quantum Information, University of Science and Technology of China, CAS, Hefei 230026, China}
\affiliation{Synergetic Innovation Center in Quantum Information and Quantum Physics, University of Science and Technology of China, CAS, Hefei 230026, China}
\author{Barry C. Sanders}
\affiliation{Synergetic Innovation Center in Quantum Information and Quantum Physics, University of Science and Technology of China, CAS, Hefei 230026, China}
\affiliation{Hefei National Laboratory for Physical Sciences at Microscale, University of Science and Technology of China, CAS, Hefei 230026, China}
\affiliation{Institute for Quantum Science and Technology, University of Calgary, Alberta T2N 1N4, Canada}
\affiliation{Program in Quantum Information Science, Canadian Institute for Advanced Research, Toronto, Ontario M5G 1M1, Canada}
\author{Wei Yi}\email{wyiz@ustc.edu.cn}
\affiliation{Key Laboratory of Quantum Information, University of Science and Technology of China, CAS, Hefei 230026, China}
\affiliation{Synergetic Innovation Center in Quantum Information and Quantum Physics, University of Science and Technology of China, CAS, Hefei 230026, China}
\author{Peng Xue}\email{gnep.eux@gmail.com}
\affiliation{Department of Physics, Southeast University, Nanjing 211189, China}
\affiliation{State Key Laboratory of Precision Spectroscopy, East China Normal University, Shanghai 200062, China}

\begin{abstract}
We report the experimental detection of bulk topological invariants in non-unitary discrete-time quantum walks with single photons. The non-unitarity of the quantum dynamics is enforced by periodically performing partial measurements on the polarization of the walker photon, which effectively introduces loss to the dynamics. The topological invariant of the non-unitary quantum walk is manifested in the quantized average displacement of the walker, which is probed by monitoring the photon loss. We confirm the topological properties of the system by observing localized edge states at the boundary of regions with different topological invariants. We further demonstrate the robustness of both the topological properties and the measurement scheme of the topological invariants against disorder.
\end{abstract}


\maketitle

{\it Introduction:---}
Topological phases, with their remarkable properties, have been under intensive study in recent years~\cite{HKrmp10,QZrmp11}. Besides condensed-matter materials such as topological insulators and superconductors, topological phenomena also emerge in synthetic systems ranging from classical light and microwaves transporting in periodically modulated media~\cite{WCJS09,Rechtsmann,K2013,HMMT13,LJS14,Cheng16,Marrucci16,silberhorn16,lewenstein16,Zeunerprl,Weimannnm,BKMM13,PBKMS15,Chong15} and phononic states in mechanical oscillators~\cite{Hubers,Khanikaevnc}, to cold atoms in optical lattices~\cite{Bloch13,ETHcoldatom14,CBG15,Weitenberg2016,BS16,Weitz16,Gadway16} and photons in discrete-time quantum walks (QWs)~\cite{KB+12}. A prominent feature of a topological phase is the emergence of topologically protected edge states, which are robust under symmetry-preserving perturbations and play a crucial role in the topological functionality of the underlying system. Through the so-called bulk-boundary correspondence, the existence and the number of these edge states are dictated by the integer-valued topological invariants, which are pivotal to the characterization of topological phases~\cite{Ryu10,Kane10}. While many experiments are concerned with the detection and characterization of topological edge states, recent progress has led to the direct measurement of topological invariants in the bulk of synthetic topological systems~\cite{Marrucci16,silberhorn16,lewenstein16,Zeunerprl,Bloch13,CBG15,Weitenberg2016,RFR+17,FRH+16}. The highly controllable parameters of these synthetic simulators offer the exciting possibility of extending the study of topological matter to regimes beyond the scope of conventional electronic systems in the condensed-matter setting.

An outstanding example here is the identification of topological phenomena in dissipative non-Hermitian systems~\cite{RL09,RLL16,ESHK,KMKO16,Zeunerprl,Weimannnm,PBKMS15}. In particular, as predicted by Rudner and Levitov~\cite{RL09,RLL16}, for a particle moving on a one-dimensional superlattice with losses on a given sublattice site in each unit cell, the average displacement of the particle is quantized, and is associated with a topological invariant defined in terms of the eigenstates of the non-Hermitian Hamiltonian. This new paradigm of topology is then experimentally confirmed in~\cite{Zeunerprl}, where the quantized mean displacement of light propagating in a lossy optical waveguide array is measured and associated with topologically non-trivial properties. While the experimental system in~\cite{Zeunerprl} features continuous-time non-Hermitian dynamics, an important question is whether topological invariants can also be defined for discrete-time non-unitary dynamics. This is particularly interesting in light of the recent demonstration of topological properties in unitary discrete-time QWs~\cite{KB+12,silberhorn16,Marrucci16,RFR+17,FRH+16}.

In this work, we report the first experimental detection of bulk topological invariants in a non-unitary discrete-time QW using single photons. To exploit the quantum nature of single photons, we entangle the photonic walker with an ancillary photon, which allows the manipulation of the dynamics in a delayed-choice fashion~\cite{JFLKK13,supp}. By periodically performing partial measurements on the photon polarization of the walker~\cite{RAA17}, we realize a non-unitary QW supporting Floquet topological phases (FTPs). While a given FTP typically features two distinct topological invariants~\cite{JiangPRL,A12,AO13}, we demonstrate that both topological invariants can be detected from the average displacement of the walker. The topological properties of the quantum dynamics are confirmed by the observation of localized edge states at the boundary between regions with distinct topological invariants. Finally, we demonstrate the robustness of the topological properties as well as the measurement scheme against disorder. Our work opens up the avenue of analyzing topological features in discrete-time quantum dynamics governed by non-unitary Floquet operators.

{\it Non-unitary discrete-time QWs:---}
We start from a split-step QW governed by the unitary Floquet operator
$U'=R(\frac{\theta_1}{2})SR(\theta_2)SR(\frac{\theta_1}{2})$. Here $R(\theta)=\one_\text{w}\otimes e^{-\text{i}\theta\sigma_y}$ is the coin operator rotating the coin state about the $y$-axis by $\theta$, $\one_\text{w}=\sum_x\ket{x}\bra{x}$, $x$ denotes the position of the walker, $\sigma_y=\text{i}(-\ket{0}\bra{1}+\ket{1}\bra{0})$ is the standard Pauli operator, and $\{\ket{0},\ket{1}\}$ are two orthogonal coin states. The position shift operator $S$ is given by $S=\sum_x\big(\ket{x-1}\bra{x}\otimes\ket{0}\bra{0}+\ket{x+1}\bra{x}\otimes\ket{1}\bra{1}\big)$.

Non-unitary time-evolution is then enforced by performing the partial measurement $M_e=\one_\text{w}\otimes\sqrt{p}\ket{-}\bra{-}$ at each time step, where $0<p\leqslant 1$ is the probability of a successful (positive) measurement and $\ket{\pm}=(\ket{0}\pm\ket{1})/\sqrt{2}$~\cite{RAA17}. When the walker is not detected, a negative measurement is applied and the state evolves as $\ket{\psi_{t}}=(\widetilde{U}')^{t}\ket{\psi_{0}}$ after $t$ steps, where $\ket{\psi_0}$ is the initial state of the walker-coin system, $\widetilde{U}'=MU'$, and $M=\one_\text{w}\otimes\big(\ket{+}\bra{+}+\sqrt{1-p}\ket{-}\bra{-}\big)$. The probability of the walker being detected at $x$ during the $t$-th step is $P_\text{th}(x,t)=\bra{\psi_{t-1}}U'^{\dagger}M^\dagger_e (\ket{x}\bra{x}\otimes\one_\text{c})M_e U'\ket{\psi_{t-1}}$, where $\one_\text{c}$ is a $2\times 2$ identity operator.

\begin{figure*}
\includegraphics[width=0.95\textwidth]{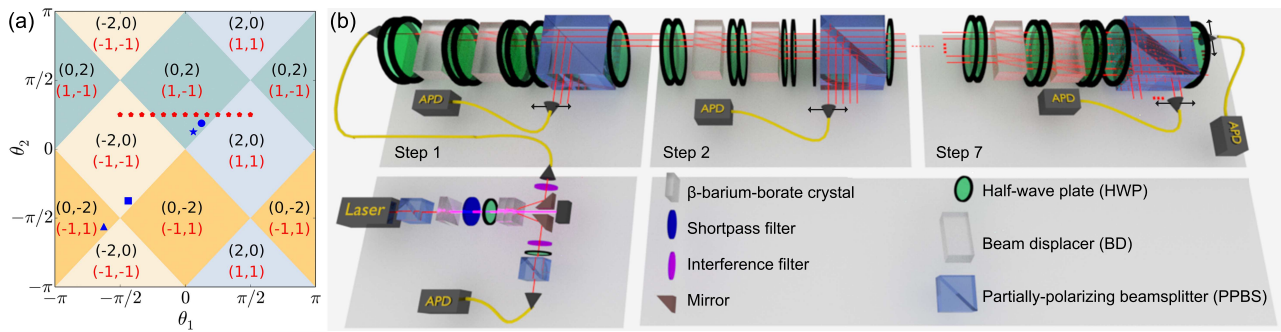}
\caption{(a) Phase diagram indicating the topological invariants $(\nu',\nu'')$ (black) and $(\nu_0,\nu_\pi)$ (red) as functions of the coin parameters $(\theta_1,\theta_2)$. The red dots indicate $13$ sets of coin parameters used to detect the topological invariants. The blue dot indicates the parameters of inhomogeneous coin rotations for the inner region and the other blue symbols for the outer regions which are used to observe the topological edge states. (b) Experimental setup. The photon pair is created via spontaneous parametric downconversion in a Bell state $\ket{\Phi^+}$. One photon as a trigger is projected in the polarization state $\ket{+}$ with a HWP and a polarizing beamsplitter. The other photon undergoes QW interferometric network. The polarization rotation $R$ and translation $S$ can be realized by two HWPs with certain setting angles and a BD, respectively. The partial measurement via loss $M_e$ is implemented by a sandwich-type HWP-PPBS-HWP setup. The photons appearing in the spatial modes at the output of a QW are coupled into optical fibers and then detected by APDs, in coincidence with the trigger photon. Photon counts are measured by translating the fiber coupler between the individual modes~\cite{BFL+10}, which gives the measured probabilities after correcting for the relative efficiencies of the different APDs.
}
\label{setup}
\end{figure*}

Interestingly, FTPs emerge in the non-unitary QW dynamics above, which can be understood by introducing the effective Hamiltonian defined through $\widetilde{U}'=\exp(-\text{i}H_{\rm eff})$. In momentum space, we have $H_{\rm eff}(k)=E_k \bm{n}\cdot\bm{\sigma}$, with $\bm{\sigma}$ the Pauli vector and $\bm{n}$ the direction of the spinor eigenstate at each momentum $-\pi<k\leq\pi$. In the limit of $p=0$, we have $M=\one_\text{w}\otimes \one_\text{c}$, and the corresponding unitary discrete-time QW features a FTP protected by chiral symmetry~\cite{CSPRA,KB+12}. The topological invariant in this case is given by the so-called winding number, i.e., the number of times the vector $\bm{n}$, which lies in the $y$-$z$ plane, winds around the $x$-axis as $k$ varies through the first Brillouin zone~\cite{CSPRA}. For $p>0$, the time-evolution becomes non-unitary, and $\bm{n}$ is no longer a real vector lying in the $y$-$z$ plane. However, as we show in the Supplemental Materials, one can still define the winding number here as the number of times the real components of $\bm{n}$ winds around the $x$-axis as $k$ varies through the first Brillouin zone~\cite{Zeunerprl,RLL16,supp}. Notably, for a given FTP, typically two distinct winding numbers $(\nu',\nu'')$ exist for Floquet operators fitted in different time frames~\cite{AO13}. While the corresponding winding number for $\widetilde{U}'$ is $\nu'$, $\nu''$ is associated with $\widetilde{U}''=MR(\frac{\theta_2}{2})SR(\theta_1)SR(\frac{\theta_2}{2})$. Alternatively, one may define the topological invariants $(\nu_0,\nu_{\pi})=[(\nu'+\nu'')/2,(\nu'-\nu'')/2]$, which are related to edge states at the boundaries through the bulk-boundary correspondence for Floquet systems~\cite{AO13,supp}. In Fig.~\ref{setup}(a), we show the phase diagram on the $\theta_1$-$\theta_2$ plane with both sets of topological invariants.

Similar to the continuous-time non-Hermitian dynamics, the topological invariants of the discrete-time non-unitary QW can be detected by measuring the average displacement of the walker initially prepared in the non-loss state $\ket{+}$ at $x=0$~\cite{RAA17}. Here the average displacement is given by $\langle \Delta x\rangle=\sum_x\sum_{t'=1}^{\infty} x P_\text{th}(x,t')$. The average dwell time, which characterizes the expected lifetime of the walker before it is measured and lost, is defined as
$\langle t \rangle=\sum_x\sum_{t'=1}^{\infty} t' P_\text{th}(x,t')$.

\begin{figure}
\includegraphics[width=0.5\textwidth]{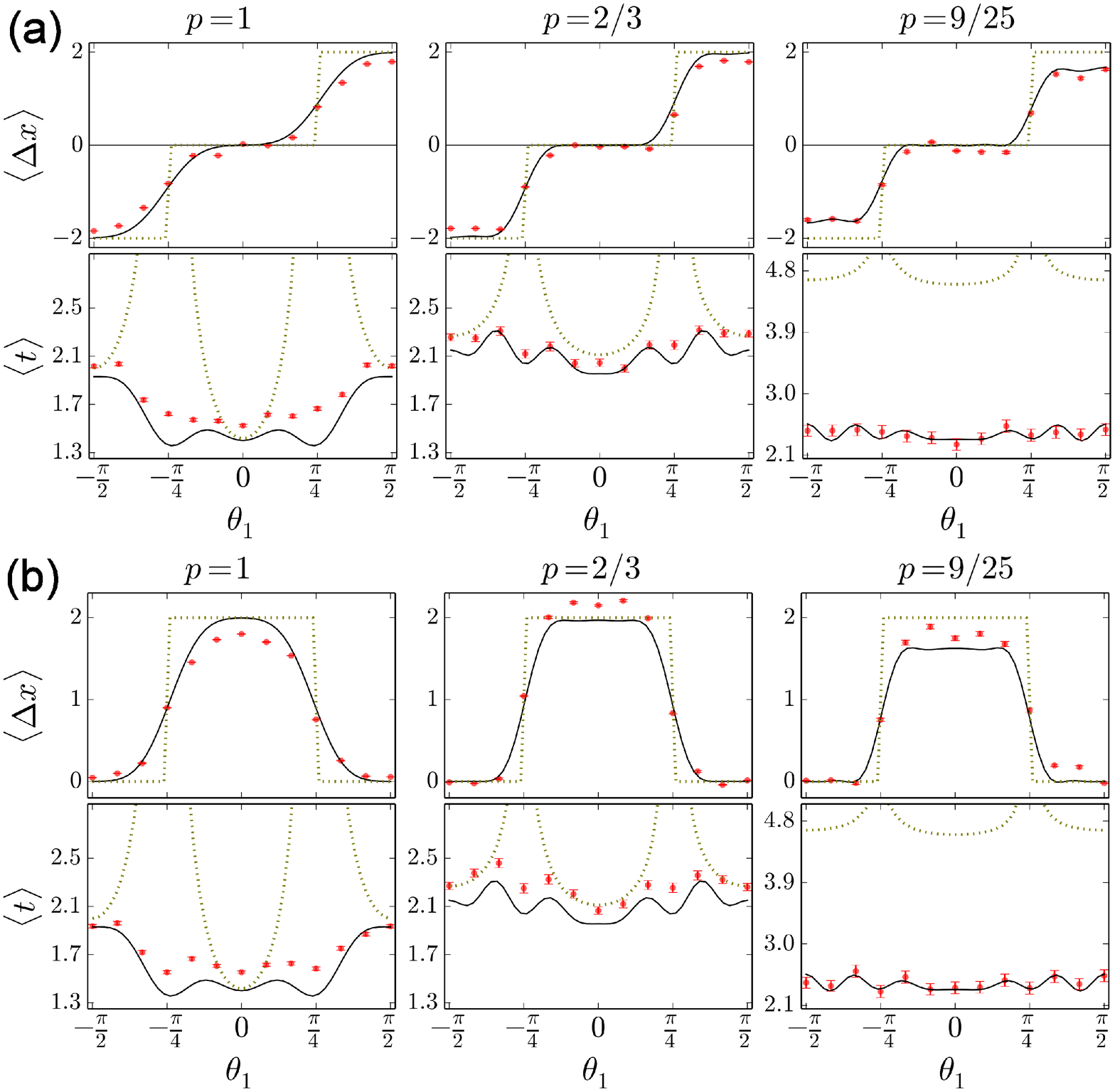}
   \caption{Measured average displacements (upper layer) and dwell time (lower layer) for $7$-step QWs of the protocols $\widetilde{U}'$ (a) and $\widetilde{U}''$ (b) as functions of the coin parameter $\theta_1$ ($\theta_2=\pi/4$ is fixed). The dashed curves indicate the expected results of infinite-step QWs~\cite{RAA17}. The solid curves indicate the numerical simulations for $7$-step QWs and the experimental results are presented by red dots. Experimental errors are due to photon counting statistics.}
\label{7step}
\end{figure}

{\it Experimental Realization of non-unitary QWs:---}
As illustrated in Fig.~\ref{setup}(b), we implement delayed-choice QWs of single photons entangled with ancillary photons, which are subject to delayed-choice coin-state projections~\cite{JFLKK13}. The coin states are represented by the horizontal $\ket{H}$ and vertical $\ket{V}$ polarization states of the photons, and the walker states are encoded in their spatial modes.

For the initial state preparation, the photon pairs are generated via type-I spontaneous parametric downconversion in a Bell state $\ket{\Phi^+}=(\ket{HH}+\ket{VV})/\sqrt{2}$. One photon serves as a trigger and is projected in $\ket{+}$ with a half-wave plate (HWP) and a polarizing beamsplitter (PBS). The other photon is delayed-chosen to be the initial coin state $\ket{+}$ heralded by the trigger photon and then sent to the QW interferometric setup. The two-photon polarization entanglement is characterized using state tomography, with the visibility $>95\%$ for  $\ket{\Phi^+}$.

The coin operator is implemented by two HWPs with certain setting angles depending on the coin parameters $(\theta_1,\theta_2)$. The shift operator $S$ is implemented via a birefringent beam displacer (BD), so that the photons in $\ket{V}$ are directly transmitted and those in $\ket{H}$ undergo a lateral displacement into a neighboring mode.

The partial measurement operator $M_e$ in each step can be realized by a sandwich-type setup involving two HWPs and a partially polarizing beamsplitter (PPBS) with the transmissivity of horizontally and vertically polarized photons $(T_\text{H},T_\text{V})=(1,1-p)$. The PPBS fully transmits horizontally polarized photons, and partially reflects vertically polarized photons, which gives the required polarization-dependent photon loss. Two HWPs at $22.5^\circ$ are used to rotate the polarization of photons (i.e., a Hadamard operation $\ket{V}\bra{-}+\ket{H}\bra{+}$ on the polarization states). Thus, photons in $\ket{-}$ are reflected by the PPBS with a probability $p$, and then detected by a single-photon avalanche photodiode (APD). The remaining photons continue the QW dynamics after another polarization rotation via the second HWP, until they are detected and lost from the system. For a $t$-step QW, we perform coincidence measurements on the reflected photons at each position successively up to $t$ and obtain $N_\text{R}(x,t')$ ($t'=1,...,t$). Together with the measurement on the number of transmitted photons $N_\text{T}(x,t)$ at the last step $t$, we construct the probability $P_\text{exp}(x,t')=N_\text{R}(x,t')/\sum_{x'}\left[\sum^{t}_{t''=1}N_\text{R}(x',t'')+N_\text{T}(x',t)\right]$, from which the average displacement and dwell time are calculated.

We note that the partial measurement operator $M_e$ is valid at the single-photon level. In our experiment, coincidence counts of up to $8000$ have been observed with the source, where heralded single photons sharing entanglement with ancillary photons are used. The probability of a multi-photon event is negligibly small ($<10^{-4}$). It is therefore reasonable to model the sandwich-type HWP-PPBS-HWP setup using $M_e$.

{\it Detecting topological invariants:---}
We realize $7$-step QWs with a fixed $\theta_2=\pi/4$ and varying $\theta_1$, for different loss parameters $p=1,2/3,9/25$.
The walker starts from $x = 0$, and the initial coin state is chosen to be $\ket{+}$ via the delayed-choice setting.
The measured average displacements are shown in Figs.~\ref{7step}(a) and (b) (upper layer) respectively for evolution operators $\widetilde{U}'$ and $\widetilde{U}''$. The results agree reasonably well with the numerical simulations of $7$-step QWs and demonstrate plateaux close to the quantized values $\nu'$ and $\nu''$ calculated from infinite-step QWs. Near topological phase transitions, where the topological invariants undergo abrupt changes, the measured average displacements deviate from the quantized topological invariants, as it takes more time for the average displacements to reach quantized values here. Such a feature can be confirmed by calculating the average dwell time of infinite-step QWs, which becomes divergent close to phase transitions~\cite{RAA17}. In Figs.~\ref{7step}(a) and (b) (lower layer), we show the measured average dwell time for $7$-step QWs, which are in a good agreement with that from numerical simulations of $7$-step QWs (solid curves). The average dwell time of finite-step QWs should approach that of infinite-step QWs (dashed curves) with increasing number of steps~\cite{RAA17,supp}. The differences between the experimental results and the simulated ones are due to experimental imperfections, especially decoherence~\cite{supp}.

\begin{figure}
\includegraphics[width=0.5\textwidth]{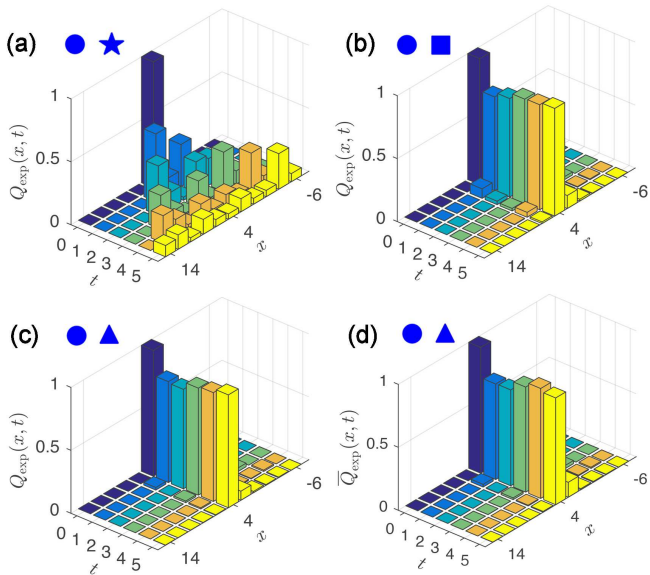}
\caption{Experimental observation of topological edge states in $5$-step inhomogeneous QWs. (a)-(c) The measured probability distributions $Q_\text{exp}(x,t)$ with fixed coin parameters for the inner region and various coin parameters for the outer regions. (d) The experimental results of the mean values of the probabilities of QWs $\overline{Q}_\text{exp}(x,t)$ with static disorder introduced to both regions.}
\label{static}
\end{figure}

{\it Topological edge states in non-unitary QWs:---}
To confirm the non-trivial topological properties of the non-unitary QW, we create regions with distinct topological invariants and probe the existence of edge states via peaked probability distribution at the boundaries for photons still evolving in the QW dynamics. For $t$-step QWs, such a probability is defined as $Q_\text{th}(x,t)=|\bra{x}\psi_t\rangle|^2/|\bra{\psi_t}\psi_t\rangle|^2$, which can be experimentally probed by normalized photon counts of the transmitted photons after step $t$ at the position $x$ via a coincidence measurement to the total number of transmitted photons, i.e., $Q_\text{exp}(x,t)=N_\text{T}(x,t)/\sum_{x'} N_\text{T}(x',t)$. The boundaries can be created by making the coin parameters $(\theta_1,\theta_2)$ spatially inhomogeneous, with $(\theta_1^\text{(o)},\theta_2^\text{(o)})$ in the outer regions ($|x|> x_0$) and $(\theta_1^\text{(i)},\theta_2^\text{(i)})$ in the inner region ($|x|\leq x_0$). These spatially inhomogeneous coin rotations in our experiment are realized via unmounted HWPs individually inserted in specific paths. We choose $x_0 = 4$ in the experiment and fix the coin parameters for the inner region as $(\theta_1^\text{(i)},\theta_2^\text{(i)})=(\pi/8,3\pi/16)$, which belong to the topological phase with bulk topological invariants $(\nu_0,\nu_{\pi})=(1,-1)$. The walker is initialized at $x = 4$ next to a boundary, and the initial coin state is chosen to be $\ket{+}$ via the delayed-choice setting. For the partial measurement, we fix $p=2/3$.

First, we choose the coin parameters $(\theta_1^\text{(o)},\theta_2^\text{(o)})=(\pi/16,\pi/8)$ for the outer regions. As the coins for both the inner and outer regions belong to the same topological phase with $(\nu_0,\nu_\pi) = (1,-1)$, no edge state is expected. As shown in Fig.~3(a), the distribution up to $5$ steps tends to extend with time, and no enhanced probability is observed near the boundary between the inner and outer regions. The wave functions of photons are distributed ballistically, similar to that of the homogenous QW. Here we define the similarity between the measured probability distribution and its theoretical prediction as a
figure of merit $S(t)=(\sum_x \sqrt{Q_\text{exp}(x,t)Q_\text{th}(x,t)})^2$, which ranges between $1$ for a perfect match and $0$ for a complete mismatch. For the first case, we obtain $S(t=5)=0.897\pm 0.018$~\cite{supp}.

Second, we choose the coin parameters $(-7\pi/16,-3\pi/8)$ for the outer regions, such that these regions belong to the topological phase with $(\nu_0,\nu_\pi)=(-1,-1)$. As the topological invariants $\nu_0$ for the inner and outer regions are different, we expect the existence of topological edge states near the boundary. As shown in Fig.~3(b), the expansion of the wave packet is highly suppressed, and the probability of the photons being detected near the boundary $Q_\text{exp}(x=4,t)$ is greatly enhanced, which confirms the existence of localized edge states. The similarity is calculated as $S(t=5)=0.988\pm 0.002$, which confirms that the experimentally measured probabilities agree well with the theoretical predictions.

Last, we change the coin parameters for the outer regions to $(-5\pi/8,-9\pi/16)$. Then both topological invariants for the outer regions $(\nu_0,\nu_\pi)=(-1,1)$ are different from those for the inner region. Our experimental results clearly show the enhanced probability $Q_\text{exp}(x = 4,t)$ in Fig.~3(c), thus confirming the existence of edge states near the boundary. Here the similarity is $S(t=5)=0.985\pm 0.002$.

\begin{figure}
\includegraphics[width=0.5\textwidth]{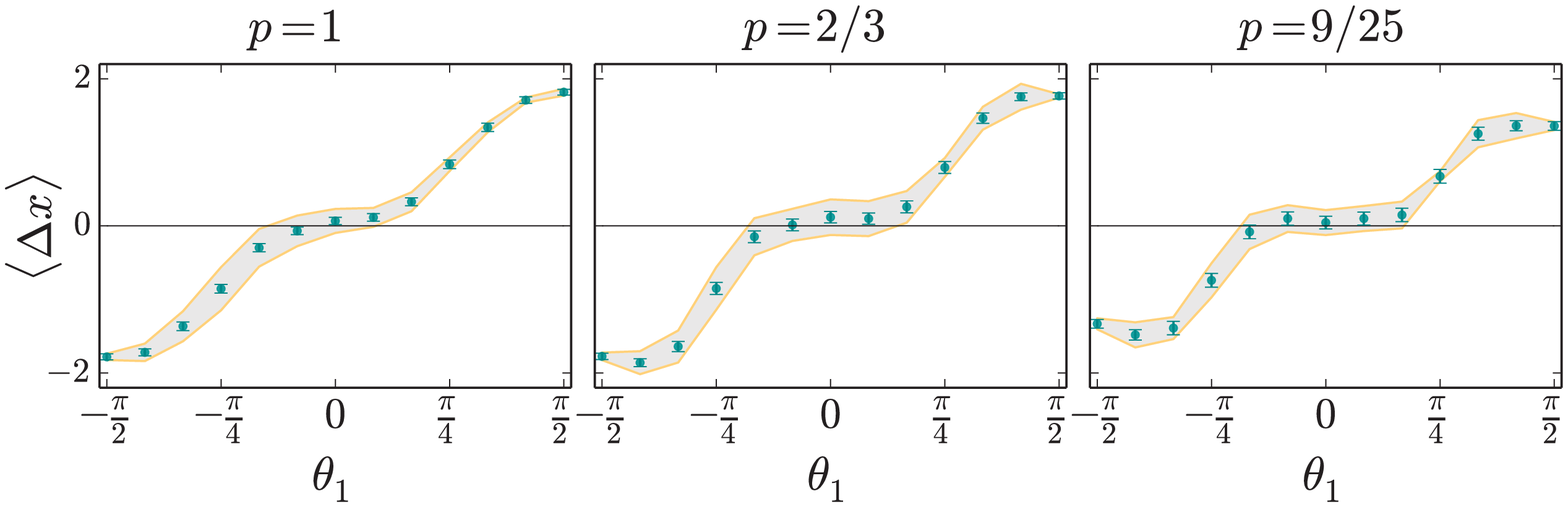}
\caption{Average displacements for $5$-step QWs with static disordered rotation angles $\theta_{1,2}+\delta\theta$, where $\delta\theta$ is unique for each position and chosen from the intervals $\left[-\pi/20,\pi/20\right]$. The coin parameters $\theta_1$ and $\theta_2$ are scanned along the dotted line in Fig.~\ref{setup}(a). The symbols and grey shadings respectively indicate the mean values of the measured average displacements and the range of the standard deviations averaged over $10$ different ensembles for each pair $(\theta_1,\theta_2)$.}
\label{static}
\end{figure}

{\it Robustness against disorder:---}
A key feature of topologically non-trivial systems is the robustness of the topological properties against small perturbations. We find that both the edge states and the quantization of the average displacement of the non-unitary QW here are robust against static disorder. We first fix $(\langle\theta_1^\text{(i)}\rangle,\langle\theta_2^\text{(i)}\rangle)=(\pi/8,3\pi/16)$ and $(\langle\theta_1^\text{(o)}\rangle,\langle\theta_2^\text{(o)}\rangle)=(-5\pi/8,-9\pi/16)$ for the inner and outer regions, respectively. Static disorder is then introduced to both regions by randomly modulating coin angles within the interval $\left[\langle\theta_{1,2}^\text{(i),(o)}\rangle-\pi/20,\langle\theta_{1,2}^\text{(i),(o)}\rangle+\pi/20\right]$. The disorder is static in time and independent at each spatial position.
We measure $P_\text{exp}(x,t')$ ($t'=1,...,5$) of $5$-step QWs with $10$ sets of randomly generated coin rotations for each position and calculate the mean values of the probabilities $\overline{Q}_\text{exp}(x,t)$. As shown in Fig.~3(d), an enhanced peak in $\overline{Q}_\text{exp}(x=4,t)$ is still observed near the boundary between the regions up to $5$ steps, with the similarity in the last step $S(t=5)=0.987\pm 0.001$.

To test the robustness of the quantization of the average displacement against static disorder, we fix the mean value $\langle \theta_2\rangle=\pi/4$ and scan $\langle\theta_1\rangle$ from $-\pi/2$ to $\pi/2$. We then measure $P_\text{exp}(x,t')$ ($t'=1,...,5$) of the $5$-step QWs governed by the evolution operator $\widetilde{U}'$ with $10$ randomly generated coin rotations for each position, and calculate the mean values of the $10$ sets of the average displacements. As shown in Fig.~4, the mean values of the average displacements are still quantized, which confirms the robustness of the measurement scheme.

{\it Conclusion:---}
We have presented the first experimental observation of bulk topological invariants for non-unitary discrete-time QWs. By choosing Floquet operators in different time frames, we are able to detect the two distinct topological invariants for the non-unitary FTP. Our work would stimulate further studies of topological phenomena in non-unitary quantum dynamics in a variety of physical systems.

\begin{acknowledgments}
This work has been supported by the Natural Science Foundation of China (Grant Nos. 11474049, 11674056, 11374283, and 11522545) and the Natural Science Foundation of Jiangsu Province (Grant No. BK20160024). WY acknowledges support from the National Key R\&D Program (Grant No. 2016YFA0301700) and the ``Strategic Priority Research Program(B)" of the Chinese Academy of Sciences (Grant No. XDB01030200). BCS acknowledges financial support from the 1000-Talent Plan.
\end{acknowledgments}

\bibliography{lossQW}

\begin{thebibliography}{42}%
\makeatletter
\providecommand \@ifxundefined [1]{%
 \@ifx{#1\undefined}
}%
\providecommand \@ifnum [1]{%
 \ifnum #1\expandafter \@firstoftwo
 \else \expandafter \@secondoftwo
 \fi
}%
\providecommand \@ifx [1]{%
 \ifx #1\expandafter \@firstoftwo
 \else \expandafter \@secondoftwo
 \fi
}%
\providecommand \natexlab [1]{#1}%
\providecommand \enquote  [1]{``#1''}%
\providecommand \bibnamefont  [1]{#1}%
\providecommand \bibfnamefont [1]{#1}%
\providecommand \citenamefont [1]{#1}%
\providecommand \href@noop [0]{\@secondoftwo}%
\providecommand \href [0]{\begingroup \@sanitize@url \@href}%
\providecommand \@href[1]{\@@startlink{#1}\@@href}%
\providecommand \@@href[1]{\endgroup#1\@@endlink}%
\providecommand \@sanitize@url [0]{\catcode `\\12\catcode `\$12\catcode
  `\&12\catcode `\#12\catcode `\^12\catcode `\_12\catcode `\%12\relax}%
\providecommand \@@startlink[1]{}%
\providecommand \@@endlink[0]{}%
\providecommand \url  [0]{\begingroup\@sanitize@url \@url }%
\providecommand \@url [1]{\endgroup\@href {#1}{\urlprefix }}%
\providecommand \urlprefix  [0]{URL }%
\providecommand \Eprint [0]{\href }%
\providecommand \doibase [0]{http://dx.doi.org/}%
\providecommand \selectlanguage [0]{\@gobble}%
\providecommand \bibinfo  [0]{\@secondoftwo}%
\providecommand \bibfield  [0]{\@secondoftwo}%
\providecommand \translation [1]{[#1]}%
\providecommand \BibitemOpen [0]{}%
\providecommand \bibitemStop [0]{}%
\providecommand \bibitemNoStop [0]{.\EOS\space}%
\providecommand \EOS [0]{\spacefactor3000\relax}%
\providecommand \BibitemShut  [1]{\csname bibitem#1\endcsname}%
\let\auto@bib@innerbib\@empty
\bibitem [{\citenamefont {Hasan}\ and\ \citenamefont {Kane}(2010)}]{HKrmp10}%
  \BibitemOpen
  \bibfield  {author} {\bibinfo {author} {\bibfnamefont {M.~Z.}\ \bibnamefont
  {Hasan}}\ and\ \bibinfo {author} {\bibfnamefont {C.~L.}\ \bibnamefont
  {Kane}},\ }\bibfield  {title} {\enquote {\bibinfo {title} {Topological
  insulators},}\ }\href {\doibase 10.1103/RevModPhys.82.3045} {\bibfield
  {journal} {\bibinfo  {journal} {Rev. Mod. Phys.}\ }\textbf {\bibinfo {volume}
  {82}},\ \bibinfo {pages} {3045--3067} (\bibinfo {year} {2010})}\BibitemShut
  {NoStop}%
\bibitem [{\citenamefont {Qi}\ and\ \citenamefont {Zhang}(2011)}]{QZrmp11}%
  \BibitemOpen
  \bibfield  {author} {\bibinfo {author} {\bibfnamefont {X.-L.}\ \bibnamefont
  {Qi}}\ and\ \bibinfo {author} {\bibfnamefont {S.-C.}\ \bibnamefont {Zhang}},\
  }\bibfield  {title} {\enquote {\bibinfo {title} {Topological insulators and
  superconductors},}\ }\href {\doibase 10.1103/RevModPhys.83.1057} {\bibfield
  {journal} {\bibinfo  {journal} {Rev. Mod. Phys.}\ }\textbf {\bibinfo {volume}
  {83}},\ \bibinfo {pages} {1057--1110} (\bibinfo {year} {2011})}\BibitemShut
  {NoStop}%
\bibitem [{\citenamefont {Wang}\ \emph {et~al.}(2009)\citenamefont {Wang},
  \citenamefont {Chong}, \citenamefont {Joannopoulos},\ and\ \citenamefont
  {Solja{\v{c}}i{\'c}}}]{WCJS09}%
  \BibitemOpen
  \bibfield  {author} {\bibinfo {author} {\bibfnamefont {Z.}~\bibnamefont
  {Wang}}, \bibinfo {author} {\bibfnamefont {Y.}~\bibnamefont {Chong}},
  \bibinfo {author} {\bibfnamefont {J.~D.}\ \bibnamefont {Joannopoulos}}, \
  and\ \bibinfo {author} {\bibfnamefont {M.}~\bibnamefont
  {Solja{\v{c}}i{\'c}}},\ }\bibfield  {title} {\enquote {\bibinfo {title}
  {Observation of unidirectional backscattering-immune topological
  electromagnetic states},}\ }\href@noop {} {\bibfield  {journal} {\bibinfo
  {journal} {Nature}\ }\textbf {\bibinfo {volume} {461}},\ \bibinfo {pages}
  {772--775} (\bibinfo {year} {2009})}\BibitemShut {NoStop}%
\bibitem [{\citenamefont {Rechtsman}\ \emph {et~al.}(2013)\citenamefont
  {Rechtsman}, \citenamefont {Zeuner}, \citenamefont {Plotnik}, \citenamefont
  {Lumer}, \citenamefont {Podolsky}, \citenamefont {Dreisow}, \citenamefont
  {Nolte}, \citenamefont {Segev},\ and\ \citenamefont {Szameit}}]{Rechtsmann}%
  \BibitemOpen
  \bibfield  {author} {\bibinfo {author} {\bibfnamefont {M.~C.}\ \bibnamefont
  {Rechtsman}}, \bibinfo {author} {\bibfnamefont {J.~M.}\ \bibnamefont
  {Zeuner}}, \bibinfo {author} {\bibfnamefont {Y.}~\bibnamefont {Plotnik}},
  \bibinfo {author} {\bibfnamefont {Y.}~\bibnamefont {Lumer}}, \bibinfo
  {author} {\bibfnamefont {D.}~\bibnamefont {Podolsky}}, \bibinfo {author}
  {\bibfnamefont {F.}~\bibnamefont {Dreisow}}, \bibinfo {author} {\bibfnamefont
  {S.}~\bibnamefont {Nolte}}, \bibinfo {author} {\bibfnamefont
  {M.}~\bibnamefont {Segev}}, \ and\ \bibinfo {author} {\bibfnamefont
  {A.}~\bibnamefont {Szameit}},\ }\bibfield  {title} {\enquote {\bibinfo
  {title} {Photonic {F}loquet topological insulators},}\ }\href@noop {}
  {\bibfield  {journal} {\bibinfo  {journal} {Nature}\ }\textbf {\bibinfo
  {volume} {496}},\ \bibinfo {pages} {196--200} (\bibinfo {year}
  {2013})}\BibitemShut {NoStop}%
\bibitem [{\citenamefont {Khanikaev}\ \emph {et~al.}(2013)\citenamefont
  {Khanikaev}, \citenamefont {Mousavi}, \citenamefont {Tse}, \citenamefont
  {Kargarian}, \citenamefont {MacDonald},\ and\ \citenamefont
  {Shvets}}]{K2013}%
  \BibitemOpen
  \bibfield  {author} {\bibinfo {author} {\bibfnamefont {A.~B.}\ \bibnamefont
  {Khanikaev}}, \bibinfo {author} {\bibfnamefont {S.~H.}\ \bibnamefont
  {Mousavi}}, \bibinfo {author} {\bibfnamefont {W.-K.}\ \bibnamefont {Tse}},
  \bibinfo {author} {\bibfnamefont {M.}~\bibnamefont {Kargarian}}, \bibinfo
  {author} {\bibfnamefont {A.~H.}\ \bibnamefont {MacDonald}}, \ and\ \bibinfo
  {author} {\bibfnamefont {G.}~\bibnamefont {Shvets}},\ }\bibfield  {title}
  {\enquote {\bibinfo {title} {Photonic topological insulators},}\ }\href@noop
  {} {\bibfield  {journal} {\bibinfo  {journal} {Nat. Mater.}\ }\textbf
  {\bibinfo {volume} {12}},\ \bibinfo {pages} {233--239} (\bibinfo {year}
  {2013})}\BibitemShut {NoStop}%
\bibitem [{\citenamefont {Hafezi}\ \emph {et~al.}(2013)\citenamefont {Hafezi},
  \citenamefont {Mittal}, \citenamefont {Fan}, \citenamefont {Migdall},\ and\
  \citenamefont {Taylor}}]{HMMT13}%
  \BibitemOpen
  \bibfield  {author} {\bibinfo {author} {\bibfnamefont {M.}~\bibnamefont
  {Hafezi}}, \bibinfo {author} {\bibfnamefont {S.}~\bibnamefont {Mittal}},
  \bibinfo {author} {\bibfnamefont {J.}~\bibnamefont {Fan}}, \bibinfo {author}
  {\bibfnamefont {A.}~\bibnamefont {Migdall}}, \ and\ \bibinfo {author}
  {\bibfnamefont {J.~M.}\ \bibnamefont {Taylor}},\ }\bibfield  {title}
  {\enquote {\bibinfo {title} {Imaging topological edge states in silicon
  photonics},}\ }\href@noop {} {\bibfield  {journal} {\bibinfo  {journal} {Nat.
  Photonics}\ }\textbf {\bibinfo {volume} {7}},\ \bibinfo {pages} {1001--1005}
  (\bibinfo {year} {2013})}\BibitemShut {NoStop}%
\bibitem [{\citenamefont {Lu}\ \emph {et~al.}(2014)\citenamefont {Lu},
  \citenamefont {Joannopoulos},\ and\ \citenamefont
  {Solja{\v{c}}i{\'c}}}]{LJS14}%
  \BibitemOpen
  \bibfield  {author} {\bibinfo {author} {\bibfnamefont {L.}~\bibnamefont
  {Lu}}, \bibinfo {author} {\bibfnamefont {J.~D.}\ \bibnamefont
  {Joannopoulos}}, \ and\ \bibinfo {author} {\bibfnamefont {M.}~\bibnamefont
  {Solja{\v{c}}i{\'c}}},\ }\bibfield  {title} {\enquote {\bibinfo {title}
  {Topological photonics},}\ }\href@noop {} {\bibfield  {journal} {\bibinfo
  {journal} {Nat. Photonics}\ }\textbf {\bibinfo {volume} {8}},\ \bibinfo
  {pages} {821--829} (\bibinfo {year} {2014})}\BibitemShut {NoStop}%
\bibitem [{\citenamefont {Cheng}\ \emph {et~al.}(2016)\citenamefont {Cheng},
  \citenamefont {Jouvaud}, \citenamefont {Ni}, \citenamefont {Mousavi},
  \citenamefont {Genack},\ and\ \citenamefont {Khanikaev}}]{Cheng16}%
  \BibitemOpen
  \bibfield  {author} {\bibinfo {author} {\bibfnamefont {X.}~\bibnamefont
  {Cheng}}, \bibinfo {author} {\bibfnamefont {C.}~\bibnamefont {Jouvaud}},
  \bibinfo {author} {\bibfnamefont {X.}~\bibnamefont {Ni}}, \bibinfo {author}
  {\bibfnamefont {S.~H.}\ \bibnamefont {Mousavi}}, \bibinfo {author}
  {\bibfnamefont {A.~Z.}\ \bibnamefont {Genack}}, \ and\ \bibinfo {author}
  {\bibfnamefont {A.~B.}\ \bibnamefont {Khanikaev}},\ }\bibfield  {title}
  {\enquote {\bibinfo {title} {Robust reconfigurable electromagnetic pathways
  within a photonic topological insulator},}\ }\href@noop {} {\bibfield
  {journal} {\bibinfo  {journal} {Nat. Mater.}\ }\textbf {\bibinfo {volume}
  {15}},\ \bibinfo {pages} {542} (\bibinfo {year} {2016})}\BibitemShut
  {NoStop}%
\bibitem [{\citenamefont {Cardano}\ \emph {et~al.}(2016)\citenamefont
  {Cardano}, \citenamefont {Maffei}, \citenamefont {Massa}, \citenamefont
  {Piccirillo}, \citenamefont {De~Lisio}, \citenamefont {De~Filippis},
  \citenamefont {Cataudella}, \citenamefont {Santamato},\ and\ \citenamefont
  {Marrucci}}]{Marrucci16}%
  \BibitemOpen
  \bibfield  {author} {\bibinfo {author} {\bibfnamefont {F.}~\bibnamefont
  {Cardano}}, \bibinfo {author} {\bibfnamefont {M.}~\bibnamefont {Maffei}},
  \bibinfo {author} {\bibfnamefont {F.}~\bibnamefont {Massa}}, \bibinfo
  {author} {\bibfnamefont {B.}~\bibnamefont {Piccirillo}}, \bibinfo {author}
  {\bibfnamefont {C.}~\bibnamefont {De~Lisio}}, \bibinfo {author}
  {\bibfnamefont {G.}~\bibnamefont {De~Filippis}}, \bibinfo {author}
  {\bibfnamefont {V.}~\bibnamefont {Cataudella}}, \bibinfo {author}
  {\bibfnamefont {E.}~\bibnamefont {Santamato}}, \ and\ \bibinfo {author}
  {\bibfnamefont {L.}~\bibnamefont {Marrucci}},\ }\bibfield  {title} {\enquote
  {\bibinfo {title} {Statistical moments of quantum-walk dynamics reveal
  topological quantum transitions},}\ }\href@noop {} {\bibfield  {journal}
  {\bibinfo  {journal} {Nat. Commun.}\ }\textbf {\bibinfo {volume} {7}},\
  \bibinfo {pages} {11439} (\bibinfo {year} {2016})}\BibitemShut {NoStop}%
\bibitem [{\citenamefont {Barkhofen}\ \emph {et~al.}()\citenamefont
  {Barkhofen}, \citenamefont {Nitsche}, \citenamefont {Elster}, \citenamefont
  {Lorz}, \citenamefont {Gabris}, \citenamefont {Jex},\ and\ \citenamefont
  {Silberhorn}}]{silberhorn16}%
  \BibitemOpen
  \bibfield  {author} {\bibinfo {author} {\bibfnamefont {S.}~\bibnamefont
  {Barkhofen}}, \bibinfo {author} {\bibfnamefont {T.}~\bibnamefont {Nitsche}},
  \bibinfo {author} {\bibfnamefont {F.}~\bibnamefont {Elster}}, \bibinfo
  {author} {\bibfnamefont {L.}~\bibnamefont {Lorz}}, \bibinfo {author}
  {\bibfnamefont {A.}~\bibnamefont {Gabris}}, \bibinfo {author} {\bibfnamefont
  {I.}~\bibnamefont {Jex}}, \ and\ \bibinfo {author} {\bibfnamefont
  {C.}~\bibnamefont {Silberhorn}},\ }\bibfield  {title} {\enquote {\bibinfo
  {title} {Measuring topological invariants and protected bound states in
  disordered discrete time quantum walks},}\ }\href@noop {} {\bibinfo
  {journal} {arXiv:1606.00299}\ }\BibitemShut {NoStop}%
\bibitem [{\citenamefont {Cardano}\ \emph {et~al.}(2017)\citenamefont
  {Cardano}, \citenamefont {D'~Errico}, \citenamefont {Dauphin}, \citenamefont
  {Maffei}, \citenamefont {Piccirillo}, \citenamefont {de~Lisio}, \citenamefont
  {De~Filippis}, \citenamefont {Cataudella}, \citenamefont {Santamato},
  \citenamefont {Marrucci}, \citenamefont {Lewenstein},\ and\ \citenamefont
  {Massignan}}]{lewenstein16}%
  \BibitemOpen
\bibfield  {journal} {  }\bibfield  {author} {\bibinfo {author} {\bibfnamefont
  {F.}~\bibnamefont {Cardano}}, \bibinfo {author} {\bibfnamefont
  {A.}~\bibnamefont {D'~Errico}}, \bibinfo {author} {\bibfnamefont
  {A.}~\bibnamefont {Dauphin}}, \bibinfo {author} {\bibfnamefont
  {M.}~\bibnamefont {Maffei}}, \bibinfo {author} {\bibfnamefont
  {B.}~\bibnamefont {Piccirillo}}, \bibinfo {author} {\bibfnamefont
  {C.}~\bibnamefont {de~Lisio}}, \bibinfo {author} {\bibfnamefont
  {G.}~\bibnamefont {De~Filippis}}, \bibinfo {author} {\bibfnamefont
  {V.}~\bibnamefont {Cataudella}}, \bibinfo {author} {\bibfnamefont
  {E.}~\bibnamefont {Santamato}}, \bibinfo {author} {\bibfnamefont
  {L.}~\bibnamefont {Marrucci}}, \bibinfo {author} {\bibfnamefont
  {M.}~\bibnamefont {Lewenstein}}, \ and\ \bibinfo {author} {\bibfnamefont
  {P.}~\bibnamefont {Massignan}},\ }\bibfield  {title} {\enquote {\bibinfo
  {title} {Detection of {Z}ak phases and topological invariants in a chiral
  quantum walk of twisted photons},}\ }\href@noop {} {\bibfield  {journal}
  {\bibinfo  {journal} {Nat. Commun.}\ }\textbf {\bibinfo {volume} {8}},\
  \bibinfo {pages} {15516} (\bibinfo {year} {2017})}\BibitemShut {NoStop}%
\bibitem [{\citenamefont {Zeuner}\ \emph {et~al.}(2015)\citenamefont {Zeuner},
  \citenamefont {Rechtsman}, \citenamefont {Plotnik}, \citenamefont {Lumer},
  \citenamefont {Nolte}, \citenamefont {Rudner}, \citenamefont {Segev},\ and\
  \citenamefont {Szameit}}]{Zeunerprl}%
  \BibitemOpen
  \bibfield  {author} {\bibinfo {author} {\bibfnamefont {J.~M.}\ \bibnamefont
  {Zeuner}}, \bibinfo {author} {\bibfnamefont {M.~C.}\ \bibnamefont
  {Rechtsman}}, \bibinfo {author} {\bibfnamefont {Y.}~\bibnamefont {Plotnik}},
  \bibinfo {author} {\bibfnamefont {Y.}~\bibnamefont {Lumer}}, \bibinfo
  {author} {\bibfnamefont {S.}~\bibnamefont {Nolte}}, \bibinfo {author}
  {\bibfnamefont {M.~S.~S.}\ \bibnamefont {Rudner}}, \bibinfo {author}
  {\bibfnamefont {M.}~\bibnamefont {Segev}}, \ and\ \bibinfo {author}
  {\bibfnamefont {A.}~\bibnamefont {Szameit}},\ }\bibfield  {title} {\enquote
  {\bibinfo {title} {Observation of a topological transition in the bulk of a
  non-{H}ermitian system},}\ }\href@noop {} {\bibfield  {journal} {\bibinfo
  {journal} {Phys. Rev. Lett.}\ }\textbf {\bibinfo {volume} {115}},\ \bibinfo
  {pages} {040402} (\bibinfo {year} {2015})}\BibitemShut {NoStop}%
\bibitem [{\citenamefont {Weimann}\ \emph {et~al.}(2017)\citenamefont
  {Weimann}, \citenamefont {Kremer}, \citenamefont {Plotnik}, \citenamefont
  {Lumer}, \citenamefont {Nolte}, \citenamefont {Makris}, \citenamefont
  {Segev}, \citenamefont {Rechtsman},\ and\ \citenamefont
  {Szameit}}]{Weimannnm}%
  \BibitemOpen
  \bibfield  {author} {\bibinfo {author} {\bibfnamefont {S.}~\bibnamefont
  {Weimann}}, \bibinfo {author} {\bibfnamefont {M.}~\bibnamefont {Kremer}},
  \bibinfo {author} {\bibfnamefont {Y.}~\bibnamefont {Plotnik}}, \bibinfo
  {author} {\bibfnamefont {Y.}~\bibnamefont {Lumer}}, \bibinfo {author}
  {\bibfnamefont {S.}~\bibnamefont {Nolte}}, \bibinfo {author} {\bibfnamefont
  {K.~G.}\ \bibnamefont {Makris}}, \bibinfo {author} {\bibfnamefont
  {M.}~\bibnamefont {Segev}}, \bibinfo {author} {\bibfnamefont {M.~C.}\
  \bibnamefont {Rechtsman}}, \ and\ \bibinfo {author} {\bibfnamefont
  {A.}~\bibnamefont {Szameit}},\ }\bibfield  {title} {\enquote {\bibinfo
  {title} {Topologically protected bound states in photonic
  parity-time-symmetric crystals},}\ }\href@noop {} {\bibfield  {journal}
  {\bibinfo  {journal} {Nat. Mater.}\ }\textbf {\bibinfo {volume} {16}},\
  \bibinfo {pages} {433--438} (\bibinfo {year} {2017})}\BibitemShut {NoStop}%
\bibitem [{\citenamefont {Bellec}\ \emph {et~al.}(2013)\citenamefont {Bellec},
  \citenamefont {Kuhl}, \citenamefont {Montambaux},\ and\ \citenamefont
  {Mortessagne}}]{BKMM13}%
  \BibitemOpen
  \bibfield  {author} {\bibinfo {author} {\bibfnamefont {M.}~\bibnamefont
  {Bellec}}, \bibinfo {author} {\bibfnamefont {U.}~\bibnamefont {Kuhl}},
  \bibinfo {author} {\bibfnamefont {G.}~\bibnamefont {Montambaux}}, \ and\
  \bibinfo {author} {\bibfnamefont {F.}~\bibnamefont {Mortessagne}},\
  }\bibfield  {title} {\enquote {\bibinfo {title} {Topological transition of
  {D}irac points in a microwave experiment},}\ }\href@noop {} {\bibfield
  {journal} {\bibinfo  {journal} {Phys. Rev. Lett.}\ }\textbf {\bibinfo
  {volume} {110}},\ \bibinfo {pages} {033902} (\bibinfo {year}
  {2013})}\BibitemShut {NoStop}%
\bibitem [{\citenamefont {Poli}\ \emph {et~al.}(2015)\citenamefont {Poli},
  \citenamefont {Bellec}, \citenamefont {Kuhl}, \citenamefont {Mortessagne},\
  and\ \citenamefont {Schomerus}}]{PBKMS15}%
  \BibitemOpen
  \bibfield  {author} {\bibinfo {author} {\bibfnamefont {C.}~\bibnamefont
  {Poli}}, \bibinfo {author} {\bibfnamefont {M.}~\bibnamefont {Bellec}},
  \bibinfo {author} {\bibfnamefont {U.}~\bibnamefont {Kuhl}}, \bibinfo {author}
  {\bibfnamefont {F.}~\bibnamefont {Mortessagne}}, \ and\ \bibinfo {author}
  {\bibfnamefont {H.}~\bibnamefont {Schomerus}},\ }\bibfield  {title} {\enquote
  {\bibinfo {title} {Selective enhancement of topologically induced interface
  states in a dielectric resonator chain},}\ }\href@noop {} {\bibfield
  {journal} {\bibinfo  {journal} {Nat. Commun.}\ }\textbf {\bibinfo {volume}
  {6}},\ \bibinfo {pages} {6710} (\bibinfo {year} {2015})}\BibitemShut
  {NoStop}%
\bibitem [{\citenamefont {Hu}\ \emph {et~al.}(2015)\citenamefont {Hu},
  \citenamefont {Pillay}, \citenamefont {Wu}, \citenamefont {Pasek},
  \citenamefont {Shum},\ and\ \citenamefont {Chong}}]{Chong15}%
  \BibitemOpen
  \bibfield  {author} {\bibinfo {author} {\bibfnamefont {W.}~\bibnamefont
  {Hu}}, \bibinfo {author} {\bibfnamefont {J.~C.}\ \bibnamefont {Pillay}},
  \bibinfo {author} {\bibfnamefont {K.}~\bibnamefont {Wu}}, \bibinfo {author}
  {\bibfnamefont {M.}~\bibnamefont {Pasek}}, \bibinfo {author} {\bibfnamefont
  {P.~P.}\ \bibnamefont {Shum}}, \ and\ \bibinfo {author} {\bibfnamefont
  {Y.~D.}\ \bibnamefont {Chong}},\ }\bibfield  {title} {\enquote {\bibinfo
  {title} {Measurement of a topological edge invariant in a microwave
  network},}\ }\href@noop {} {\bibfield  {journal} {\bibinfo  {journal} {Phys.
  Rev. X}\ }\textbf {\bibinfo {volume} {5}},\ \bibinfo {pages} {011012}
  (\bibinfo {year} {2015})}\BibitemShut {NoStop}%
\bibitem [{\citenamefont {S{\"u}sstrunk}\ and\ \citenamefont
  {Huber}(2015)}]{Hubers}%
  \BibitemOpen
  \bibfield  {author} {\bibinfo {author} {\bibfnamefont {R.}~\bibnamefont
  {S{\"u}sstrunk}}\ and\ \bibinfo {author} {\bibfnamefont {S.~D.}\ \bibnamefont
  {Huber}},\ }\bibfield  {title} {\enquote {\bibinfo {title} {Observation of
  phononic helical edge states in a mechanical topological insulator},}\
  }\href@noop {} {\bibfield  {journal} {\bibinfo  {journal} {Science}\ }\textbf
  {\bibinfo {volume} {349}},\ \bibinfo {pages} {47--50} (\bibinfo {year}
  {2015})}\BibitemShut {NoStop}%
\bibitem [{\citenamefont {Fleury}\ \emph {et~al.}(2016)\citenamefont {Fleury},
  \citenamefont {Khanikaev},\ and\ \citenamefont {Al{\`u}}}]{Khanikaevnc}%
  \BibitemOpen
  \bibfield  {author} {\bibinfo {author} {\bibfnamefont {R.}~\bibnamefont
  {Fleury}}, \bibinfo {author} {\bibfnamefont {A.~B.}\ \bibnamefont
  {Khanikaev}}, \ and\ \bibinfo {author} {\bibfnamefont {A.}~\bibnamefont
  {Al{\`u}}},\ }\bibfield  {title} {\enquote {\bibinfo {title} {{F}loquet
  topological insulators for sound},}\ }\href@noop {} {\bibfield  {journal}
  {\bibinfo  {journal} {Nat. Commun.}\ }\textbf {\bibinfo {volume} {7}},\
  \bibinfo {pages} {11744} (\bibinfo {year} {2016})}\BibitemShut {NoStop}%
\bibitem [{\citenamefont {Atala}\ \emph {et~al.}(2013)\citenamefont {Atala},
  \citenamefont {Aidelsburger}, \citenamefont {Barreiro}, \citenamefont
  {Abanin}, \citenamefont {Kitagawa}, \citenamefont {Demler},\ and\
  \citenamefont {Bloch}}]{Bloch13}%
  \BibitemOpen
  \bibfield  {author} {\bibinfo {author} {\bibfnamefont {M.}~\bibnamefont
  {Atala}}, \bibinfo {author} {\bibfnamefont {M.}~\bibnamefont {Aidelsburger}},
  \bibinfo {author} {\bibfnamefont {J.~T.}\ \bibnamefont {Barreiro}}, \bibinfo
  {author} {\bibfnamefont {D.}~\bibnamefont {Abanin}}, \bibinfo {author}
  {\bibfnamefont {T.}~\bibnamefont {Kitagawa}}, \bibinfo {author}
  {\bibfnamefont {E.}~\bibnamefont {Demler}}, \ and\ \bibinfo {author}
  {\bibfnamefont {I.}~\bibnamefont {Bloch}},\ }\bibfield  {title} {\enquote
  {\bibinfo {title} {Direct measurement of the {Z}ak phase in topological
  {B}loch bands},}\ }\href@noop {} {\bibfield  {journal} {\bibinfo  {journal}
  {Nat. Phys.}\ }\textbf {\bibinfo {volume} {9}},\ \bibinfo {pages} {795--800}
  (\bibinfo {year} {2013})}\BibitemShut {NoStop}%
\bibitem [{\citenamefont {Jotzu}\ \emph {et~al.}(2014)\citenamefont {Jotzu},
  \citenamefont {Messer}, \citenamefont {Desbuquois}, \citenamefont {Lebrat},
  \citenamefont {Uehlinger}, \citenamefont {Greif},\ and\ \citenamefont
  {Esslinger}}]{ETHcoldatom14}%
  \BibitemOpen
  \bibfield  {author} {\bibinfo {author} {\bibfnamefont {G.}~\bibnamefont
  {Jotzu}}, \bibinfo {author} {\bibfnamefont {M.}~\bibnamefont {Messer}},
  \bibinfo {author} {\bibfnamefont {R.}~\bibnamefont {Desbuquois}}, \bibinfo
  {author} {\bibfnamefont {M.}~\bibnamefont {Lebrat}}, \bibinfo {author}
  {\bibfnamefont {T.}~\bibnamefont {Uehlinger}}, \bibinfo {author}
  {\bibfnamefont {D.}~\bibnamefont {Greif}}, \ and\ \bibinfo {author}
  {\bibfnamefont {T.}~\bibnamefont {Esslinger}},\ }\bibfield  {title} {\enquote
  {\bibinfo {title} {Experimental realization of the topological {H}aldane
  model with ultracold fermions},}\ }\href@noop {} {\bibfield  {journal}
  {\bibinfo  {journal} {Nature}\ }\textbf {\bibinfo {volume} {515}},\ \bibinfo
  {pages} {237--240} (\bibinfo {year} {2014})}\BibitemShut {NoStop}%
\bibitem [{\citenamefont {Aidelsburger}\ \emph {et~al.}(2015)\citenamefont
  {Aidelsburger}, \citenamefont {Lohse}, \citenamefont {Schweizer},
  \citenamefont {Atala}, \citenamefont {Barreiro}, \citenamefont {Nascimbene},
  \citenamefont {Cooper}, \citenamefont {Bloch},\ and\ \citenamefont
  {Goldman}}]{CBG15}%
  \BibitemOpen
  \bibfield  {author} {\bibinfo {author} {\bibfnamefont {M.}~\bibnamefont
  {Aidelsburger}}, \bibinfo {author} {\bibfnamefont {M.}~\bibnamefont {Lohse}},
  \bibinfo {author} {\bibfnamefont {C.}~\bibnamefont {Schweizer}}, \bibinfo
  {author} {\bibfnamefont {M.}~\bibnamefont {Atala}}, \bibinfo {author}
  {\bibfnamefont {J.~T.}\ \bibnamefont {Barreiro}}, \bibinfo {author}
  {\bibfnamefont {S.}~\bibnamefont {Nascimbene}}, \bibinfo {author}
  {\bibfnamefont {N.~R.}\ \bibnamefont {Cooper}}, \bibinfo {author}
  {\bibfnamefont {I.}~\bibnamefont {Bloch}}, \ and\ \bibinfo {author}
  {\bibfnamefont {N.}~\bibnamefont {Goldman}},\ }\bibfield  {title} {\enquote
  {\bibinfo {title} {Measuring the {C}hern number of {H}ofstadter bands with
  ultracold bosonic atoms},}\ }\href@noop {} {\bibfield  {journal} {\bibinfo
  {journal} {Nat. Phys.}\ }\textbf {\bibinfo {volume} {11}},\ \bibinfo {pages}
  {162--166} (\bibinfo {year} {2015})}\BibitemShut {NoStop}%
\bibitem [{\citenamefont {Fl{\"a}schner}\ \emph {et~al.}(2016)\citenamefont
  {Fl{\"a}schner}, \citenamefont {Rem}, \citenamefont {Tarnowski},
  \citenamefont {Vogel}, \citenamefont {L{\"u}hmann}, \citenamefont
  {Sengstock},\ and\ \citenamefont {Weitenberg}}]{Weitenberg2016}%
  \BibitemOpen
  \bibfield  {author} {\bibinfo {author} {\bibfnamefont {N.}~\bibnamefont
  {Fl{\"a}schner}}, \bibinfo {author} {\bibfnamefont {B.~S.}\ \bibnamefont
  {Rem}}, \bibinfo {author} {\bibfnamefont {M.}~\bibnamefont {Tarnowski}},
  \bibinfo {author} {\bibfnamefont {D.}~\bibnamefont {Vogel}}, \bibinfo
  {author} {\bibfnamefont {D.-S.}\ \bibnamefont {L{\"u}hmann}}, \bibinfo
  {author} {\bibfnamefont {K.}~\bibnamefont {Sengstock}}, \ and\ \bibinfo
  {author} {\bibfnamefont {C.}~\bibnamefont {Weitenberg}},\ }\bibfield  {title}
  {\enquote {\bibinfo {title} {Experimental reconstruction of the {B}erry
  curvature in a {F}loquet {B}loch band},}\ }\href@noop {} {\bibfield
  {journal} {\bibinfo  {journal} {Science}\ }\textbf {\bibinfo {volume}
  {352}},\ \bibinfo {pages} {1091--1094} (\bibinfo {year} {2016})}\BibitemShut
  {NoStop}%
\bibitem [{\citenamefont {Li}\ \emph {et~al.}(2016)\citenamefont {Li},
  \citenamefont {Duca}, \citenamefont {Reitter}, \citenamefont {Grusdt},
  \citenamefont {Demler}, \citenamefont {Endres}, \citenamefont
  {Schleier-Smith}, \citenamefont {Bloch},\ and\ \citenamefont
  {Schneider}}]{BS16}%
  \BibitemOpen
  \bibfield  {author} {\bibinfo {author} {\bibfnamefont {T.}~\bibnamefont
  {Li}}, \bibinfo {author} {\bibfnamefont {L.}~\bibnamefont {Duca}}, \bibinfo
  {author} {\bibfnamefont {M.}~\bibnamefont {Reitter}}, \bibinfo {author}
  {\bibfnamefont {F.}~\bibnamefont {Grusdt}}, \bibinfo {author} {\bibfnamefont
  {E.}~\bibnamefont {Demler}}, \bibinfo {author} {\bibfnamefont
  {M.}~\bibnamefont {Endres}}, \bibinfo {author} {\bibfnamefont
  {M.}~\bibnamefont {Schleier-Smith}}, \bibinfo {author} {\bibfnamefont
  {I.}~\bibnamefont {Bloch}}, \ and\ \bibinfo {author} {\bibfnamefont
  {U.}~\bibnamefont {Schneider}},\ }\bibfield  {title} {\enquote {\bibinfo
  {title} {{B}loch state tomography using {W}ilson lines},}\ }\href@noop {}
  {\bibfield  {journal} {\bibinfo  {journal} {Science}\ }\textbf {\bibinfo
  {volume} {352}},\ \bibinfo {pages} {1094--1097} (\bibinfo {year}
  {2016})}\BibitemShut {NoStop}%
\bibitem [{\citenamefont {Leder}\ \emph {et~al.}(2016)\citenamefont {Leder},
  \citenamefont {Grossert}, \citenamefont {Sitta}, \citenamefont {Genske},
  \citenamefont {Rosch},\ and\ \citenamefont {Weitz}}]{Weitz16}%
  \BibitemOpen
  \bibfield  {author} {\bibinfo {author} {\bibfnamefont {M.}~\bibnamefont
  {Leder}}, \bibinfo {author} {\bibfnamefont {C.}~\bibnamefont {Grossert}},
  \bibinfo {author} {\bibfnamefont {L.}~\bibnamefont {Sitta}}, \bibinfo
  {author} {\bibfnamefont {M.}~\bibnamefont {Genske}}, \bibinfo {author}
  {\bibfnamefont {A.}~\bibnamefont {Rosch}}, \ and\ \bibinfo {author}
  {\bibfnamefont {M.}~\bibnamefont {Weitz}},\ }\bibfield  {title} {\enquote
  {\bibinfo {title} {Real-space imaging of a topologically protected edge state
  with ultracold atoms in an amplitude-chirped optical lattice},}\ }\href@noop
  {} {\bibfield  {journal} {\bibinfo  {journal} {Nat. Commun.}\ }\textbf
  {\bibinfo {volume} {7}},\ \bibinfo {pages} {13112} (\bibinfo {year}
  {2016})}\BibitemShut {NoStop}%
\bibitem [{\citenamefont {Meier}\ \emph {et~al.}(2016)\citenamefont {Meier},
  \citenamefont {An},\ and\ \citenamefont {Gadway}}]{Gadway16}%
  \BibitemOpen
  \bibfield  {author} {\bibinfo {author} {\bibfnamefont {E.~J.}\ \bibnamefont
  {Meier}}, \bibinfo {author} {\bibfnamefont {F.~A.}\ \bibnamefont {An}}, \
  and\ \bibinfo {author} {\bibfnamefont {B.}~\bibnamefont {Gadway}},\
  }\bibfield  {title} {\enquote {\bibinfo {title} {Observation of the
  topological soliton state in the {S}u--{S}chrieffer--{H}eeger model},}\
  }\href@noop {} {\bibfield  {journal} {\bibinfo  {journal} {Nat. Commun.}\
  }\textbf {\bibinfo {volume} {7}},\ \bibinfo {pages} {13986} (\bibinfo {year}
  {2016})}\BibitemShut {NoStop}%
\bibitem [{\citenamefont {Kitagawa}\ \emph {et~al.}(2012)\citenamefont
  {Kitagawa}, \citenamefont {Broome}, \citenamefont {Fedrizzi}, \citenamefont
  {Rudner}, \citenamefont {Berg}, \citenamefont {Kassal}, \citenamefont
  {Aspuru-Guzik}, \citenamefont {Demler},\ and\ \citenamefont {White}}]{KB+12}%
  \BibitemOpen
  \bibfield  {author} {\bibinfo {author} {\bibfnamefont {T.}~\bibnamefont
  {Kitagawa}}, \bibinfo {author} {\bibfnamefont {M.~A.}\ \bibnamefont
  {Broome}}, \bibinfo {author} {\bibfnamefont {A.}~\bibnamefont {Fedrizzi}},
  \bibinfo {author} {\bibfnamefont {M.~S.}\ \bibnamefont {Rudner}}, \bibinfo
  {author} {\bibfnamefont {E.}~\bibnamefont {Berg}}, \bibinfo {author}
  {\bibfnamefont {I.}~\bibnamefont {Kassal}}, \bibinfo {author} {\bibfnamefont
  {A.}~\bibnamefont {Aspuru-Guzik}}, \bibinfo {author} {\bibfnamefont
  {E.}~\bibnamefont {Demler}}, \ and\ \bibinfo {author} {\bibfnamefont {A.~G.}\
  \bibnamefont {White}},\ }\bibfield  {title} {\enquote {\bibinfo {title}
  {Observation of topologically protected bound states in photonic quantum
  walks},}\ }\href@noop {} {\bibfield  {journal} {\bibinfo  {journal} {Nat.
  Commun.}\ }\textbf {\bibinfo {volume} {3}},\ \bibinfo {pages} {882} (\bibinfo
  {year} {2012})}\BibitemShut {NoStop}%
\bibitem [{\citenamefont {Ryu}\ \emph {et~al.}(2010)\citenamefont {Ryu},
  \citenamefont {Schnyder}, \citenamefont {Furusaki},\ and\ \citenamefont
  {Ludwig}}]{Ryu10}%
  \BibitemOpen
  \bibfield  {author} {\bibinfo {author} {\bibfnamefont {S.}~\bibnamefont
  {Ryu}}, \bibinfo {author} {\bibfnamefont {A.~P.}\ \bibnamefont {Schnyder}},
  \bibinfo {author} {\bibfnamefont {A.}~\bibnamefont {Furusaki}}, \ and\
  \bibinfo {author} {\bibfnamefont {A.~W.~W.}\ \bibnamefont {Ludwig}},\
  }\bibfield  {title} {\enquote {\bibinfo {title} {Topological insulators and
  superconductors: tenfold way and dimensional hierarchy},}\ }\href@noop {}
  {\bibfield  {journal} {\bibinfo  {journal} {New J. Phys.}\ }\textbf {\bibinfo
  {volume} {12}},\ \bibinfo {pages} {065010} (\bibinfo {year}
  {2010})}\BibitemShut {NoStop}%
\bibitem [{\citenamefont {Teo}\ and\ \citenamefont {Kane}(2010)}]{Kane10}%
  \BibitemOpen
  \bibfield  {author} {\bibinfo {author} {\bibfnamefont {J.~C.~Y.}\
  \bibnamefont {Teo}}\ and\ \bibinfo {author} {\bibfnamefont {C.~L.}\
  \bibnamefont {Kane}},\ }\bibfield  {title} {\enquote {\bibinfo {title}
  {Topological defects and gapless modes in insulators and superconductors},}\
  }\href@noop {} {\bibfield  {journal} {\bibinfo  {journal} {Phys. Rev. B}\
  }\textbf {\bibinfo {volume} {82}},\ \bibinfo {pages} {115120} (\bibinfo
  {year} {2010})}\BibitemShut {NoStop}%
\bibitem [{\citenamefont {Ramasesh}\ \emph {et~al.}(2017)\citenamefont
  {Ramasesh}, \citenamefont {Flurin}, \citenamefont {Rudner}, \citenamefont
  {Siddiqi},\ and\ \citenamefont {Yao}}]{RFR+17}%
  \BibitemOpen
  \bibfield  {author} {\bibinfo {author} {\bibfnamefont {V.~V.}\ \bibnamefont
  {Ramasesh}}, \bibinfo {author} {\bibfnamefont {E.}~\bibnamefont {Flurin}},
  \bibinfo {author} {\bibfnamefont {M.}~\bibnamefont {Rudner}}, \bibinfo
  {author} {\bibfnamefont {I.}~\bibnamefont {Siddiqi}}, \ and\ \bibinfo
  {author} {\bibfnamefont {N.~Y.}\ \bibnamefont {Yao}},\ }\bibfield  {title}
  {\enquote {\bibinfo {title} {Direct probe of topological invariants using
  {B}loch oscillating quantum walks},}\ }\href@noop {} {\bibfield  {journal}
  {\bibinfo  {journal} {Phys. Rev. Lett.}\ }\textbf {\bibinfo {volume} {118}},\
  \bibinfo {pages} {130501} (\bibinfo {year} {2017})}\BibitemShut {NoStop}%
\bibitem [{\citenamefont {Flurin}\ \emph {et~al.}(2017)\citenamefont {Flurin},
  \citenamefont {Ramasesh}, \citenamefont {Hacohen-Gourgy}, \citenamefont
  {Martin}, \citenamefont {Yao},\ and\ \citenamefont {Siddiqi}}]{FRH+16}%
  \BibitemOpen
  \bibfield  {author} {\bibinfo {author} {\bibfnamefont {E.}~\bibnamefont
  {Flurin}}, \bibinfo {author} {\bibfnamefont {V.~V.}\ \bibnamefont
  {Ramasesh}}, \bibinfo {author} {\bibfnamefont {S.}~\bibnamefont
  {Hacohen-Gourgy}}, \bibinfo {author} {\bibfnamefont {L.~S.}\ \bibnamefont
  {Martin}}, \bibinfo {author} {\bibfnamefont {N.~Y.}\ \bibnamefont {Yao}}, \
  and\ \bibinfo {author} {\bibfnamefont {I.}~\bibnamefont {Siddiqi}},\
  }\bibfield  {title} {\enquote {\bibinfo {title} {Observing topological
  invariants using quantum walks in superconducting circuits},}\ }\href
  {\doibase 10.1103/PhysRevX.7.031023} {\bibfield  {journal} {\bibinfo
  {journal} {Phys. Rev. X}\ }\textbf {\bibinfo {volume} {7}},\ \bibinfo {pages}
  {031023} (\bibinfo {year} {2017})}\BibitemShut {NoStop}%
\bibitem [{\citenamefont {Rudner}\ and\ \citenamefont {Levitov}(2009)}]{RL09}%
  \BibitemOpen
  \bibfield  {author} {\bibinfo {author} {\bibfnamefont {M.~S.}\ \bibnamefont
  {Rudner}}\ and\ \bibinfo {author} {\bibfnamefont {L.~S.}\ \bibnamefont
  {Levitov}},\ }\bibfield  {title} {\enquote {\bibinfo {title} {Topological
  transition in a non-{H}ermitian quantum walk},}\ }\href@noop {} {\bibfield
  {journal} {\bibinfo  {journal} {Phys. Rev. Lett.}\ }\textbf {\bibinfo
  {volume} {102}},\ \bibinfo {pages} {065703} (\bibinfo {year}
  {2009})}\BibitemShut {NoStop}%
\bibitem [{\citenamefont {Rudner}\ \emph {et~al.}(2016)\citenamefont {Rudner},
  \citenamefont {Levin},\ and\ \citenamefont {Levitov}}]{RLL16}%
  \BibitemOpen
  \bibfield  {author} {\bibinfo {author} {\bibfnamefont {M.~S.}\ \bibnamefont
  {Rudner}}, \bibinfo {author} {\bibfnamefont {M.}~\bibnamefont {Levin}}, \
  and\ \bibinfo {author} {\bibfnamefont {L.~S.}\ \bibnamefont {Levitov}},\
  }\bibfield  {title} {\enquote {\bibinfo {title} {Survival, decay, and
  topological protection in non-{H}ermitian quantum transport},}\ }\href@noop
  {} {\bibfield  {journal} {\bibinfo  {journal} {arXiv:1605.07652}\ } (\bibinfo
  {year} {2016})}\BibitemShut {NoStop}%
\bibitem [{\citenamefont {Esaki}\ \emph {et~al.}(2011)\citenamefont {Esaki},
  \citenamefont {Sato}, \citenamefont {Hasebe},\ and\ \citenamefont
  {Kohmoto}}]{ESHK}%
  \BibitemOpen
  \bibfield  {author} {\bibinfo {author} {\bibfnamefont {K.}~\bibnamefont
  {Esaki}}, \bibinfo {author} {\bibfnamefont {M.}~\bibnamefont {Sato}},
  \bibinfo {author} {\bibfnamefont {K.}~\bibnamefont {Hasebe}}, \ and\ \bibinfo
  {author} {\bibfnamefont {M.}~\bibnamefont {Kohmoto}},\ }\bibfield  {title}
  {\enquote {\bibinfo {title} {Edge states and topological phases in
  non-{H}ermitian systems},}\ }\href@noop {} {\bibfield  {journal} {\bibinfo
  {journal} {Phys. Rev. B}\ }\textbf {\bibinfo {volume} {84}},\ \bibinfo
  {pages} {205128} (\bibinfo {year} {2011})}\BibitemShut {NoStop}%
\bibitem [{\citenamefont {Kim}\ \emph {et~al.}(2016)\citenamefont {Kim},
  \citenamefont {Ken}, \citenamefont {Kawakami},\ and\ \citenamefont
  {Obuse}}]{KMKO16}%
  \BibitemOpen
  \bibfield  {author} {\bibinfo {author} {\bibfnamefont {D.}~\bibnamefont
  {Kim}}, \bibinfo {author} {\bibfnamefont {M.}~\bibnamefont {Ken}}, \bibinfo
  {author} {\bibfnamefont {N.}~\bibnamefont {Kawakami}}, \ and\ \bibinfo
  {author} {\bibfnamefont {H.}~\bibnamefont {Obuse}},\ }\bibfield  {title}
  {\enquote {\bibinfo {title} {{F}loquet topological phases driven by
  $\mathcal{PT}$ symmetric nonunitary time evolution},}\ }\href@noop {}
  {\bibfield  {journal} {\bibinfo  {journal} {arXiv:1609.09650}\ } (\bibinfo
  {year} {2016})}\BibitemShut {NoStop}%
\bibitem [{\citenamefont {Jeong}\ \emph {et~al.}(2013)\citenamefont {Jeong},
  \citenamefont {Di~Franco}, \citenamefont {Lim}, \citenamefont {Kim},\ and\
  \citenamefont {Kim}}]{JFLKK13}%
  \BibitemOpen
  \bibfield  {author} {\bibinfo {author} {\bibfnamefont {Y.-C.}\ \bibnamefont
  {Jeong}}, \bibinfo {author} {\bibfnamefont {C.}~\bibnamefont {Di~Franco}},
  \bibinfo {author} {\bibfnamefont {H.-T.}\ \bibnamefont {Lim}}, \bibinfo
  {author} {\bibfnamefont {M.~S.}\ \bibnamefont {Kim}}, \ and\ \bibinfo
  {author} {\bibfnamefont {Y.-H.}\ \bibnamefont {Kim}},\ }\bibfield  {title}
  {\enquote {\bibinfo {title} {Experimental realization of a delayed-choice
  quantum walk},}\ }\href@noop {} {\bibfield  {journal} {\bibinfo  {journal}
  {Nat. Commun.}\ }\textbf {\bibinfo {volume} {4}},\ \bibinfo {pages} {2471}
  (\bibinfo {year} {2013})}\BibitemShut {NoStop}%
\bibitem [{sup()}]{supp}%
  \BibitemOpen
  \href@noop {} {}\bibinfo {note} {See the Supplemental Material at
  http://link.aps.org/ supplemental/10.1103/PhysRevLett.119.130501 for details
  on the derivation of the topological invariants of the nonunitary
  discrete-time QW, the application of the delayed-choice QW, the convergence
  of average dwell time, the robustness of the topological properties of the QW
  and the measurement scheme of the topological invariants against dynamic
  disorder, and the impact of decoherence on the experimental
  measurements.}\BibitemShut {Stop}%
\bibitem [{\citenamefont {Rakovszky}\ \emph {et~al.}(2017)\citenamefont
  {Rakovszky}, \citenamefont {Asb{\'o}th},\ and\ \citenamefont
  {Alberti}}]{RAA17}%
  \BibitemOpen
  \bibfield  {author} {\bibinfo {author} {\bibfnamefont {T.}~\bibnamefont
  {Rakovszky}}, \bibinfo {author} {\bibfnamefont {J.~K.}\ \bibnamefont
  {Asb{\'o}th}}, \ and\ \bibinfo {author} {\bibfnamefont {A.}~\bibnamefont
  {Alberti}},\ }\bibfield  {title} {\enquote {\bibinfo {title} {Detecting
  topological invariants in chiral symmetric insulators via losses},}\
  }\href@noop {} {\bibfield  {journal} {\bibinfo  {journal} {Phys. Rev. B}\
  }\textbf {\bibinfo {volume} {95}},\ \bibinfo {pages} {201407} (\bibinfo
  {year} {2017})}\BibitemShut {NoStop}%
\bibitem [{\citenamefont {Jiang}\ \emph {et~al.}(2011)\citenamefont {Jiang},
  \citenamefont {Kitagawa}, \citenamefont {Alicea}, \citenamefont {Akhmerov},
  \citenamefont {Pekker}, \citenamefont {Refael}, \citenamefont {Cirac},
  \citenamefont {Demler}, \citenamefont {Lukin},\ and\ \citenamefont
  {Zoller}}]{JiangPRL}%
  \BibitemOpen
  \bibfield  {author} {\bibinfo {author} {\bibfnamefont {L.}~\bibnamefont
  {Jiang}}, \bibinfo {author} {\bibfnamefont {T.}~\bibnamefont {Kitagawa}},
  \bibinfo {author} {\bibfnamefont {J.}~\bibnamefont {Alicea}}, \bibinfo
  {author} {\bibfnamefont {A.~R.}\ \bibnamefont {Akhmerov}}, \bibinfo {author}
  {\bibfnamefont {D.}~\bibnamefont {Pekker}}, \bibinfo {author} {\bibfnamefont
  {G.}~\bibnamefont {Refael}}, \bibinfo {author} {\bibfnamefont {J.~I.}\
  \bibnamefont {Cirac}}, \bibinfo {author} {\bibfnamefont {E.}~\bibnamefont
  {Demler}}, \bibinfo {author} {\bibfnamefont {M.~D.}\ \bibnamefont {Lukin}}, \
  and\ \bibinfo {author} {\bibfnamefont {P.}~\bibnamefont {Zoller}},\
  }\bibfield  {title} {\enquote {\bibinfo {title} {{M}ajorana fermions in
  equilibrium and in driven cold-atom quantum wires},}\ }\href {\doibase
  10.1103/PhysRevLett.106.220402} {\bibfield  {journal} {\bibinfo  {journal}
  {Phys. Rev. Lett.}\ }\textbf {\bibinfo {volume} {106}},\ \bibinfo {pages}
  {220402} (\bibinfo {year} {2011})}\BibitemShut {NoStop}%
\bibitem [{\citenamefont {Asb\'oth}(2012)}]{A12}%
  \BibitemOpen
  \bibfield  {author} {\bibinfo {author} {\bibfnamefont {J.~K.}\ \bibnamefont
  {Asb\'oth}},\ }\bibfield  {title} {\enquote {\bibinfo {title} {Symmetries,
  topological phases, and bound states in the one-dimensional quantum walk},}\
  }\href {\doibase 10.1103/PhysRevB.86.195414} {\bibfield  {journal} {\bibinfo
  {journal} {Phys. Rev. B}\ }\textbf {\bibinfo {volume} {86}},\ \bibinfo
  {pages} {195414} (\bibinfo {year} {2012})}\BibitemShut {NoStop}%
\bibitem [{\citenamefont {Asb{\'o}th}\ and\ \citenamefont
  {Obuse}(2013)}]{AO13}%
  \BibitemOpen
  \bibfield  {author} {\bibinfo {author} {\bibfnamefont {J.~K.}\ \bibnamefont
  {Asb{\'o}th}}\ and\ \bibinfo {author} {\bibfnamefont {H.}~\bibnamefont
  {Obuse}},\ }\bibfield  {title} {\enquote {\bibinfo {title} {Bulk-boundary
  correspondence for chiral symmetric quantum walks},}\ }\href@noop {}
  {\bibfield  {journal} {\bibinfo  {journal} {Phys. Rev. B}\ }\textbf {\bibinfo
  {volume} {88}},\ \bibinfo {pages} {121406} (\bibinfo {year}
  {2013})}\BibitemShut {NoStop}%
\bibitem [{\citenamefont {Kitagawa}\ \emph {et~al.}(2010)\citenamefont
  {Kitagawa}, \citenamefont {Rudner}, \citenamefont {Berg},\ and\ \citenamefont
  {Demler}}]{CSPRA}%
  \BibitemOpen
  \bibfield  {author} {\bibinfo {author} {\bibfnamefont {T.}~\bibnamefont
  {Kitagawa}}, \bibinfo {author} {\bibfnamefont {M.~S.}\ \bibnamefont
  {Rudner}}, \bibinfo {author} {\bibfnamefont {E.}~\bibnamefont {Berg}}, \ and\
  \bibinfo {author} {\bibfnamefont {E.}~\bibnamefont {Demler}},\ }\bibfield
  {title} {\enquote {\bibinfo {title} {Exploring topological phases with
  quantum walks},}\ }\href {\doibase 10.1103/PhysRevA.82.033429} {\bibfield
  {journal} {\bibinfo  {journal} {Phys. Rev. A}\ }\textbf {\bibinfo {volume}
  {82}},\ \bibinfo {pages} {033429} (\bibinfo {year} {2010})}\BibitemShut
  {NoStop}%
\bibitem [{\citenamefont {Broome}\ \emph {et~al.}(2010)\citenamefont {Broome},
  \citenamefont {Fedrizzi}, \citenamefont {Lanyon}, \citenamefont {Kassal},
  \citenamefont {Aspuru-Guzik},\ and\ \citenamefont {White}}]{BFL+10}%
  \BibitemOpen
  \bibfield  {author} {\bibinfo {author} {\bibfnamefont {M.~A.}\ \bibnamefont
  {Broome}}, \bibinfo {author} {\bibfnamefont {A.}~\bibnamefont {Fedrizzi}},
  \bibinfo {author} {\bibfnamefont {B.~P.}\ \bibnamefont {Lanyon}}, \bibinfo
  {author} {\bibfnamefont {I.}~\bibnamefont {Kassal}}, \bibinfo {author}
  {\bibfnamefont {A.}~\bibnamefont {Aspuru-Guzik}}, \ and\ \bibinfo {author}
  {\bibfnamefont {A.~G.}\ \bibnamefont {White}},\ }\bibfield  {title} {\enquote
  {\bibinfo {title} {Discrete single-photon quantum walks with tunable
  decoherence},}\ }\href {\doibase 10.1103/PhysRevLett.104.153602} {\bibfield
  {journal} {\bibinfo  {journal} {Phys. Rev. Lett.}\ }\textbf {\bibinfo
  {volume} {104}},\ \bibinfo {pages} {153602} (\bibinfo {year}
  {2010})}\BibitemShut {NoStop}%
\end{thebibliography}%

\end{document}